\documentclass[aps,pre]{revtex4}
\usepackage{hyperref,natbib,doi}
\usepackage{graphicx}
\usepackage{amsmath}
\usepackage{dcolumn}
\usepackage[utf8]{inputenc}
\usepackage{amssymb}

\begin{document}

\title{The effect of longitudinal density gradient on electron plasma wake field acceleration}

\author{David Tsiklauri}
\affiliation{School of Physics and Astronomy, Queen Mary University of London, London, E1 4NS, United Kingdom}

\begin{abstract} 
Three dimensional, particle-in-cell, fully electromagnetic simulations of 
electron plasma wake field acceleration in the blow out regime are presented. 
Earlier results are extended by  (i) studying the effect of longitudinal density gradient;
(ii) avoiding use of co-moving simulation box; (iii) inclusion of ion motion; and (iv) studying fully 
electromagnetic plasma wake fields.
It is established that injecting driving and trailing electron bunches
into a positive density gradient of ten-fold increasing density over 10 cm long Lithium vapor plasma,
results in spatially more compact and three times larger,
compared to the
uniform density case, electric fields ($-6.4 \times 10^{10}$ V/m), 
leading to acceleration of the trailing bunch up to 24.4 GeV (starting from initial 20.4 GeV),
with an energy transfer efficiencies from leading to trailing bunch of 75 percent.
In  the
uniform density case $-2.5 \times 10^{10}$ V/m wake is created
leading to acceleration of the trailing bunch up to 22.4 GeV,
with an energy transfer efficiencies of 65 percent.
It is also established that injecting the electron bunches
into a negative density gradient of ten-fold decreasing density over 10 cm long plasma,
results in spatially more spread and two-and-half smaller electric fields ($-1.0 \times 10^{10}$ V/m), 
leading to a weaker acceleration of the trailing bunch up to 21.4 GeV,
with an energy transfer efficiencies of 45 percent.
Inclusion of ion motions into consideration shows that in the plasma wake ion number density can
increase over few times the background value.
It is also shown that transverse electromagnetic fields in plasma wake are of the same
order as the longitudinal (electrostatic) ones. 
\end{abstract}

\maketitle

\section{Introduction}

Conventional particle accelerators have an accelerating gradient of tens of MV/m. 
The limit is set by the radio frequency (RF) breakdown phenomenon, when 
large electric field in the accelerator cavities causes accelerator to be 
effectively short-circuited. It is known that when electric field exceeds a threshold 
value, the Kilpatrick limit, no matter how good the vacuum in the accelerator 
tube is, there will be a RF breakdown and nearly all of the RF 
power is absorbed \cite{k57}. Thus the beam receives little or no 
acceleration in the cavity. Novel accelerator concepts that have a 
promise to overcome the difficulties of conventional particle 
accelerators include plasma wake field acceleration (PWFA), amongst a 
number of other concepts such as fixed-field alternating-gradient 
accelerators, dielectric wall accelerators, and dielectric laser 
accelerators. Plasma wake field acceleration usually refers to an 
acceleration of particles through the generation of strong electric 
fields from charged particle motion in plasma. Generally a distinction 
is drawn between beam-driven PWFA and laser-driven  
plasma wake field acceleration (LWFA). The advantage of accelerators based on 
plasmas, is that they can sustain electric fields up to tens of GV/m, without 
electric short-circuiting, and have a 
potential to be a smaller and cheaper than the conventional accelerators. 
An electrostatic wave in plasma, co-propagating with a charged-particle 
bunch/beam, can keep the bunch in a high-field region over a path of a 
meter or more, thus transferring substantial amounts of energy to the 
particles in a compact space.  However, the precision engineering is 
required to accelerate particles efficiently and uniformly and this 
has been a challenge.

Recently authors of  Ref.\cite{l14}
have taken a leap forward in PWFA. In their plasma wake field 
accelerator, the plasma wave is created by a 20-GeV electron 
bunch from SLAC's linac. A second bunch of equally energetic 
electrons follows close behind. With SLAC's purpose-built 
Facility for Advanced Accelerator Experimental Tests (FACET) \cite{facet}, 
authors could place the trailing bunch at just the right spot in the plasma wave to 
increase the bunch energy by 1.6 GeV over just 30 cm of plasma. 
In Ref.\cite{l14} 3D particle-in-cell (PIC) simulations with resolution of 
$512^3$ spatial grids of a plasma wake field interaction 
with beam were also carried out.
They established that the drive bunch clears away the 
plasma electrons, leaving a region of strong but inhomogeneous electric 
field in its wake. If the trailing bunch is large enough and positioned in the 
right spot, it can flatten the electric field so that the trailing bunch is uniformly 
accelerated \cite{l14}. Thus, PWFA is an attractive concept to achieve 
the acceleration of about 74 pC of charge contained in the core of the 
trailing bunch in an accelerating gradient of about 4.4 GV/m. These core 
particles gain about 1.6 GeV of energy per particle, with a final energy 
spread as low as 0.7 \% (2.0 \% on average), and an energy-transfer 
efficiency from the wake to the bunch that can exceed 30 \% (17.7 \% on average). 
This acceleration of a distinct bunch of electrons containing a substantial 
charge and having a small energy spread with both high accelerating 
gradient and high energy-transfer efficiency represents a milestone 
in the development of PWFA into a
compact and affordable accelerator technology. 
Despite these ground breaking advances there is a room for improvements: 
(i) An energy gain by the core electron bunch of about 10\% can 
potentially be improved. 
(ii) From the 800 pC that started out in the trailing bunch, only 74 pC 
remained in the accelerated core. Better preservation of the beam is a 
priority for future work.
(iii) Previous numerical simulations of PWFA have predicted the hosing 
instability to be disrupting the efficient acceleration. This has not been 
observed in the experiments \cite{l14,facet,m15}. Further work is needed to settle this issue.

Recently an interesting opportunity has been explored by Ref.\cite{pt14}. 
In this work a beam of accelerated electrons was injected into a magnetized, 
Maxwellian, homogeneous, and inhomogeneous background plasma. It was 
established that in the case of increasing density along the path of an electron 
beam wave-particle resonant interaction of Langmuir waves (the same type of wave as in PWFA) 
with the beam electrons leads to an efficient particle acceleration.  
This is because Langmuir waves drift to smaller wave-numbers, $k$, allowing them to 
increase their phase speed, $V_{ph}=\omega/k$, and, therefore, 
being subject to absorption by faster electrons. This is a novel aspect and has not been 
yet explored in the PWFA context. Therefore the main motivation for this study is to 
explore the effect of longitudinal density gradient on 
electron plasma wake field acceleration.

Section II presents the model and results. Section III summaries the main findings.

\section{The model and results}

\begin{figure*}
\begin{center}
\includegraphics[width=16.5cm]{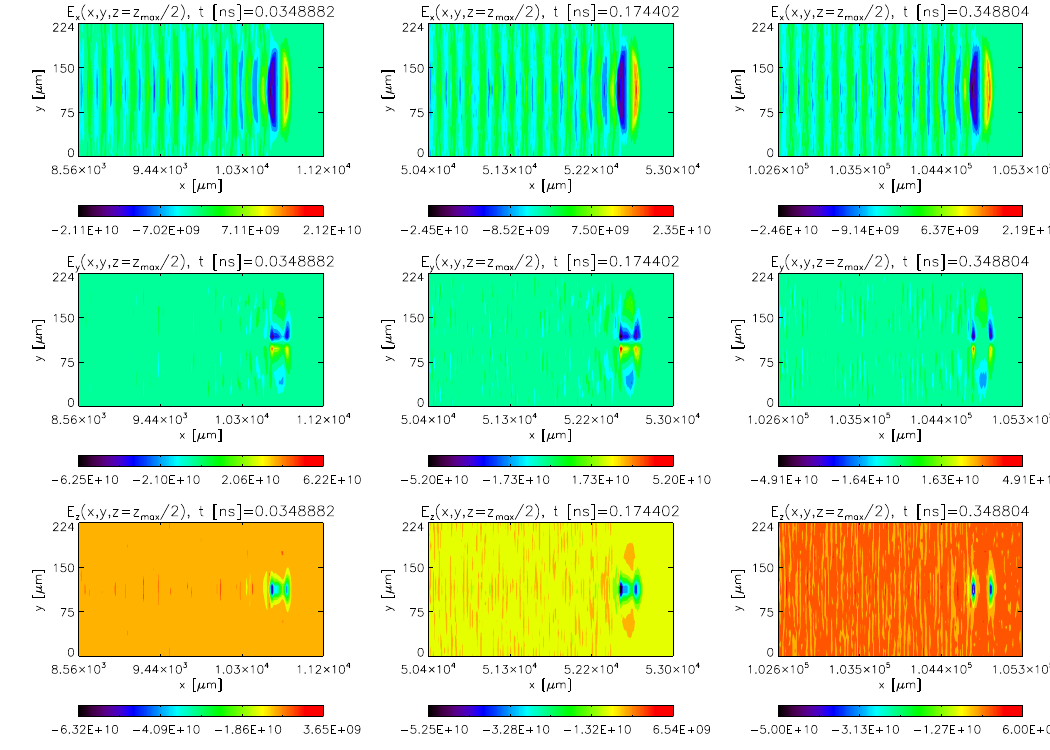}
\end{center}
\caption{Contour plots of electric field x- (top row), y- (middle row) 
and z- (bottom row) components in (x,y) plane (cut through $z=z_{max}/2$) 
at different time instants
corresponding to 1/10th, half and the final simulations times.
The fields on color bars are quoted in $V/m$ and time at the top of each panel is
in nano-seconds. The data is for uniform density run.}
\label{fig1}
\end{figure*}

\begin{figure*}
\begin{center}
\includegraphics[width=16.5cm]{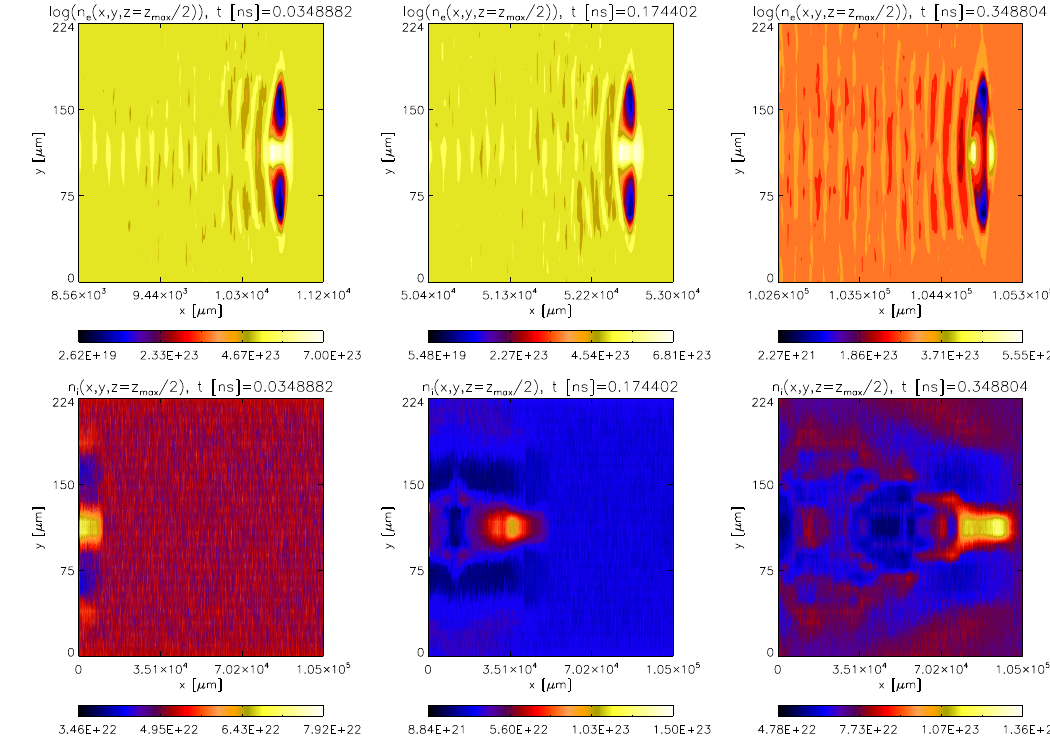}
\end{center}
\caption{Contour plots of logarithm of electron (top row)  
and ion (bottom row) in (x,y) plane (cut through $z=z_{max}/2$) 
at different time instants
corresponding to 1/10th, half and the final simulations times.
The number densities on color bars are quoted in m$^{-3}$ 
and time at the top of each panel is
in nano-seconds. The data is for uniform density run.}
\label{fig2}
\end{figure*}

\begin{figure}
\includegraphics[width=13.0cm]{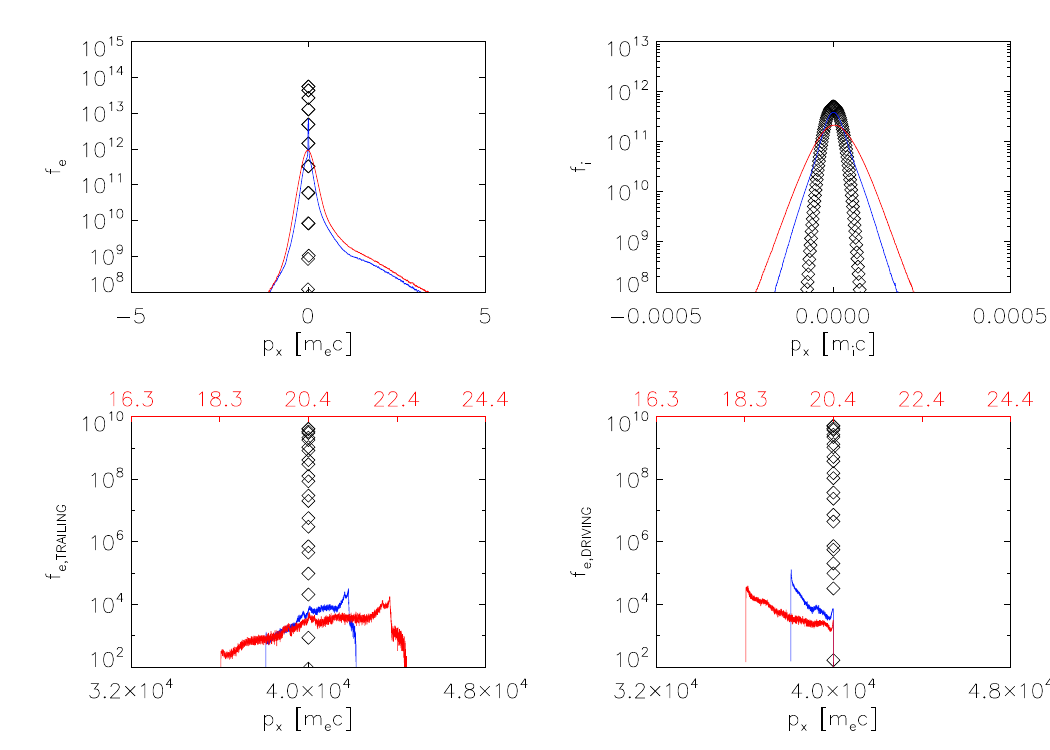}
\caption{Background electron (top-left), ion (top-right), trailing (bottom-left) and 
driving (bottom-right) electron bunch
distribution functions at different times:
open diamonds correspond to $t=0$, while blue and red curves to the 
half and the final simulations times, respectively.
x-axis are momenta quoted in the units of relevant species mass times
speed of light i.e. $[m_e c]$ or $[m_i c]$ as shown on each panel. 
In the two bottom panels, at the top the energy is quoted in GeV with red numbers, to aid
eye visualizing of
trailing bunch acceleration and driving bunch deceleration processes.
The data is for uniform density run.}
\label{fig3}
\end{figure}

\begin{figure} 
\begin{center}
\includegraphics[width=8.5cm]{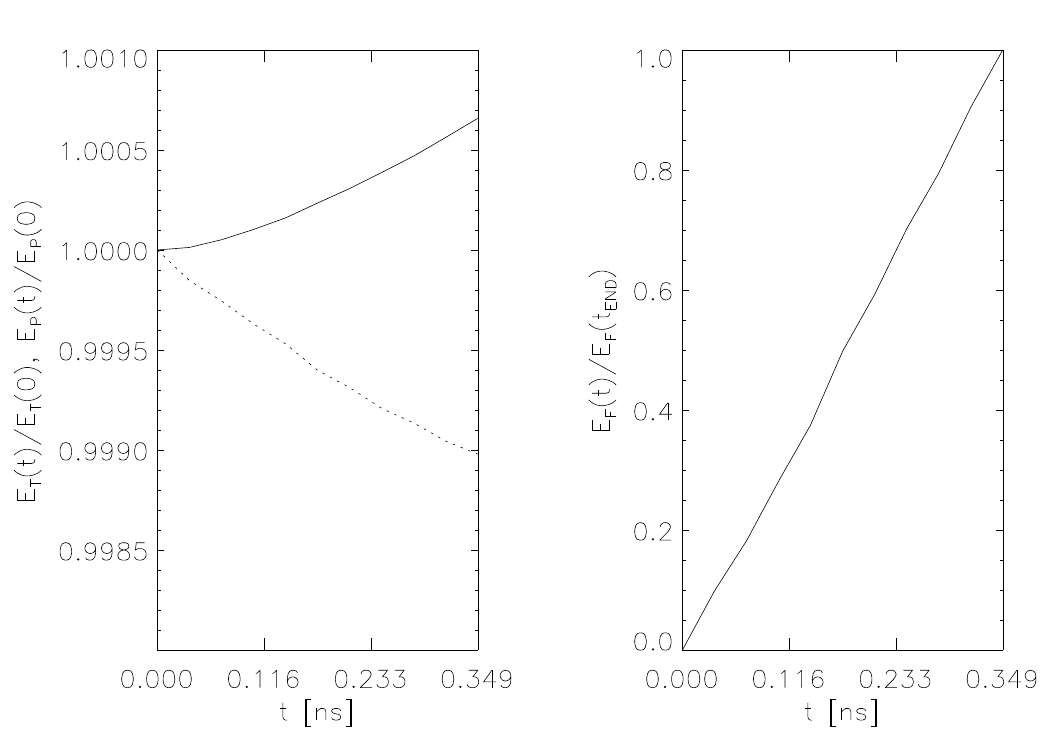}
\end{center}
\caption{Left panel's solid and dashed curves
are the total (particles plus EM fields) and 
particle energies, normalized on initial values, respectively. 
Right panel shows
EM field energy normalized on its final simulation time
value (because it is zero at $t=0$). The data is for uniform density run.}
\label{fig4}
\end{figure}

\begin{figure*}
\begin{center}
\includegraphics[width=16.5cm]{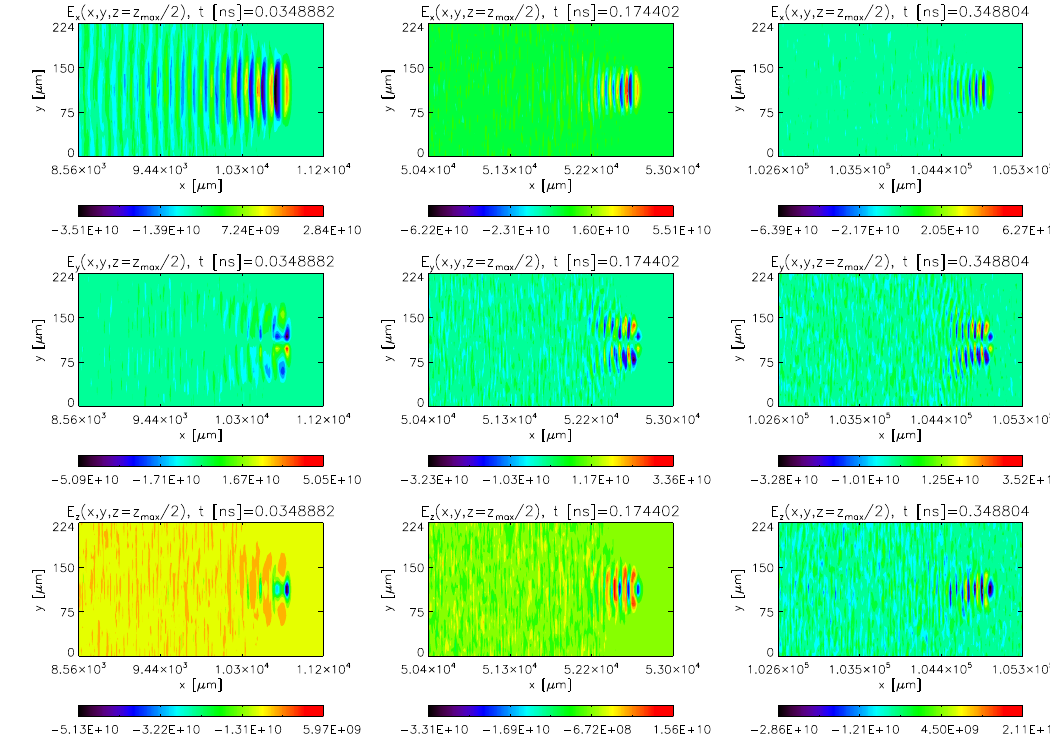}
\end{center}
\caption{As in Fig.\ref{fig1} but for the case of positive density gradient,
according to equation \ref{e1}.}
\label{fig5}
\end{figure*}

\begin{figure*}
\begin{center}
\includegraphics[width=16.5cm]{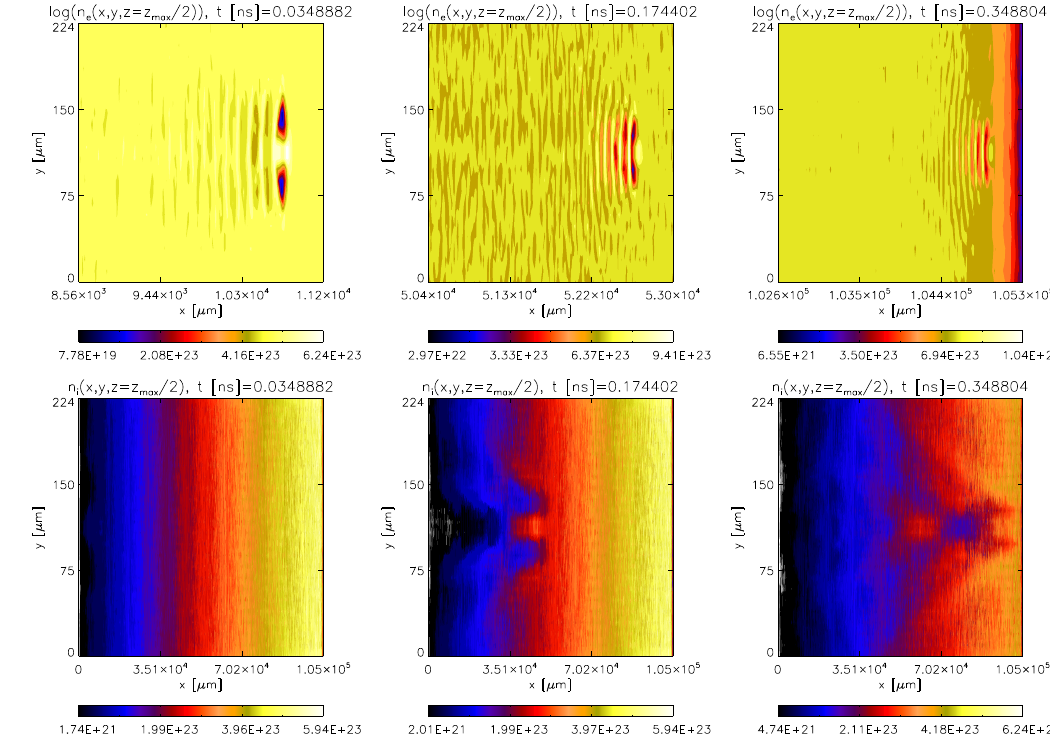}
\end{center}
\caption{As in Fig.\ref{fig2} but for the case of positive density gradient,
according to equation \ref{e1}.}
\label{fig6}
\end{figure*}

\begin{figure}
\includegraphics[width=13.0cm]{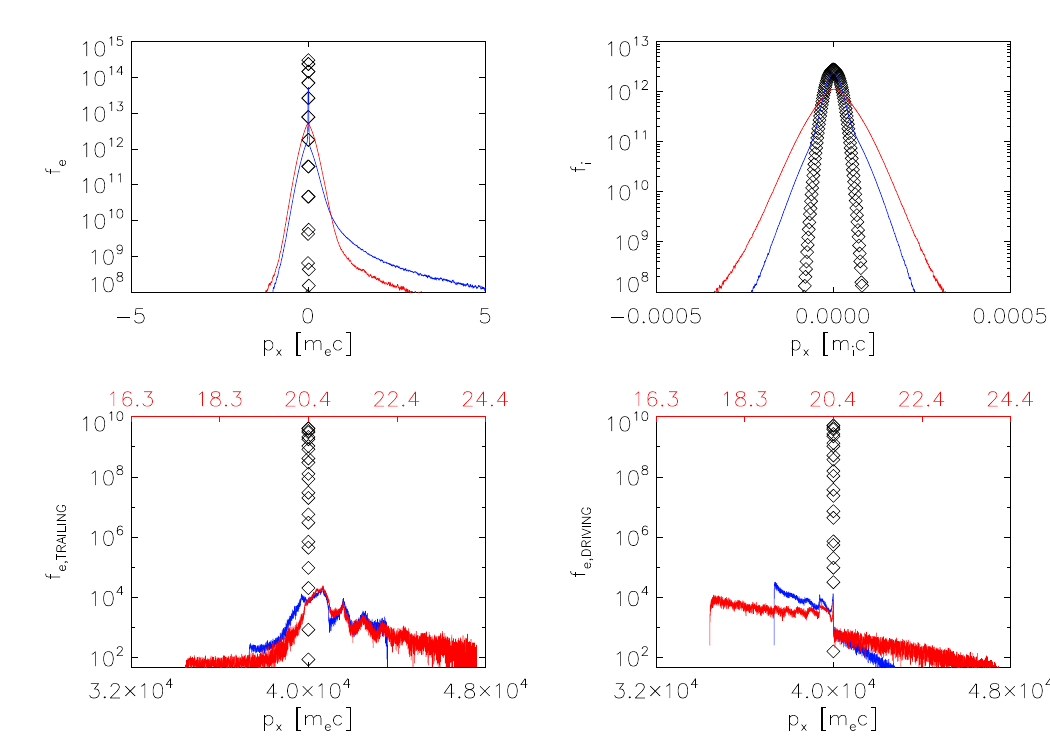}
\caption{As in Fig.\ref{fig3} but for the case of positive density gradient,
according to equation \ref{e1}.}
\label{fig7}
\end{figure}

\begin{figure} 
\begin{center}
\includegraphics[width=8.5cm]{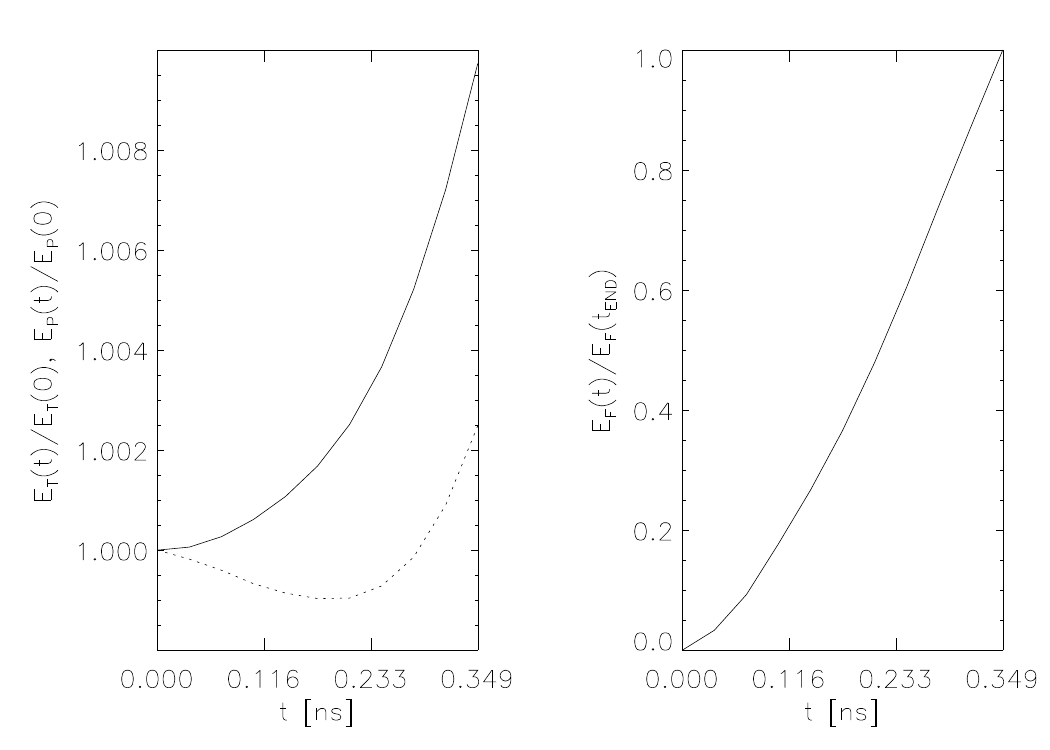}
\end{center}
\caption{As in Fig.\ref{fig4} but for the case of positive density gradient,
according to equation \ref{e1}. }
\label{fig8}
\end{figure}

\begin{figure*}
\begin{center}
\includegraphics[width=16.5cm]{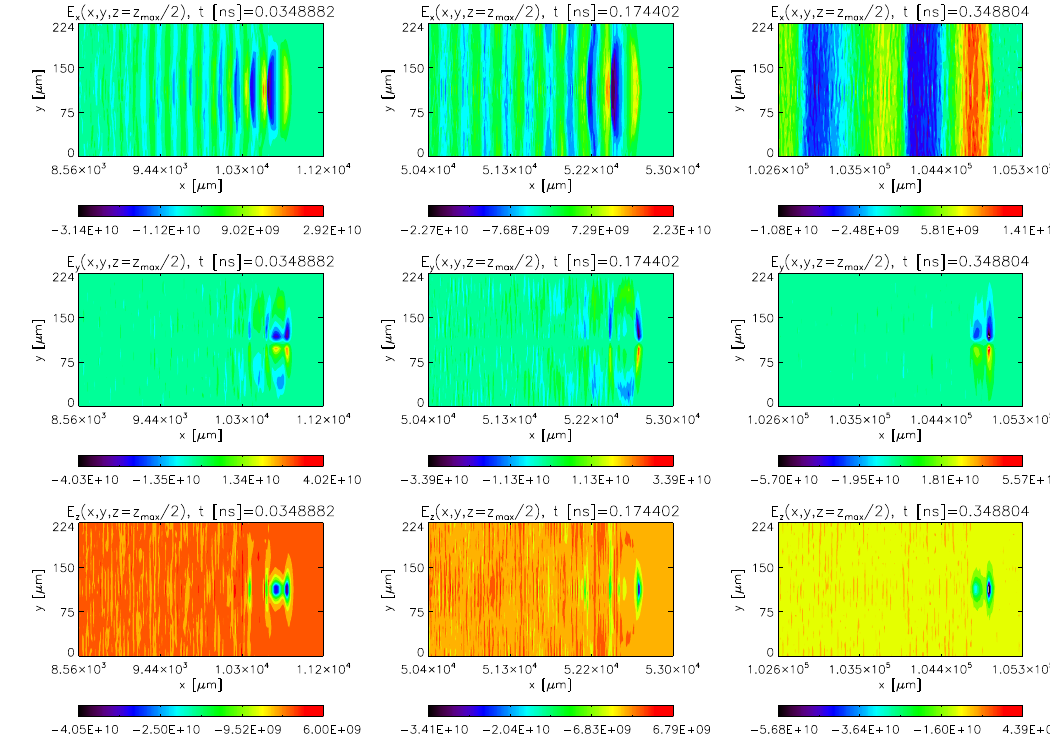}
\end{center}
\caption{As in Fig.\ref{fig1} but for the case of negative density gradient,
according to equation \ref{e2}.}
\label{fig9}
\end{figure*}

\begin{figure*}
\begin{center}
\includegraphics[width=16.5cm]{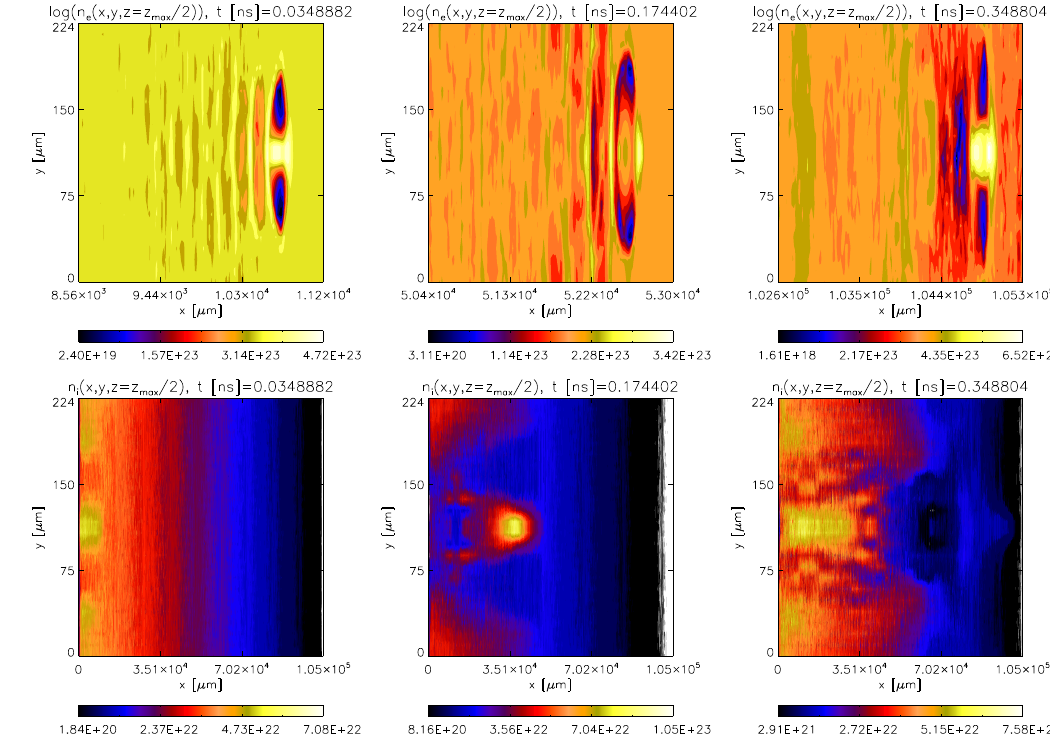}
\end{center}
\caption{As in Fig.\ref{fig2} but for the case of negative density gradient,
according to equation \ref{e2}.}
\label{fig10}
\end{figure*}

\begin{figure}
\includegraphics[width=13.0cm]{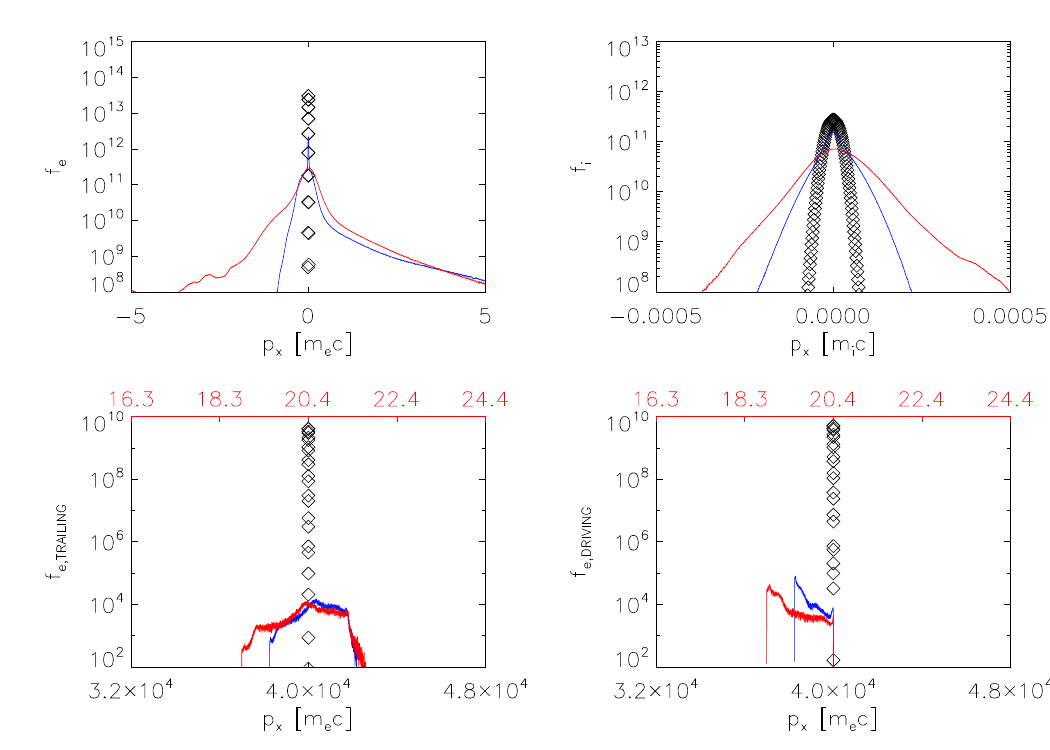}
\caption{As in Fig.\ref{fig3} but for the case of negative density gradient,
according to equation \ref{e2}.}
\label{fig11}
\end{figure}

\begin{figure} 
\begin{center}
\includegraphics[width=8.5cm]{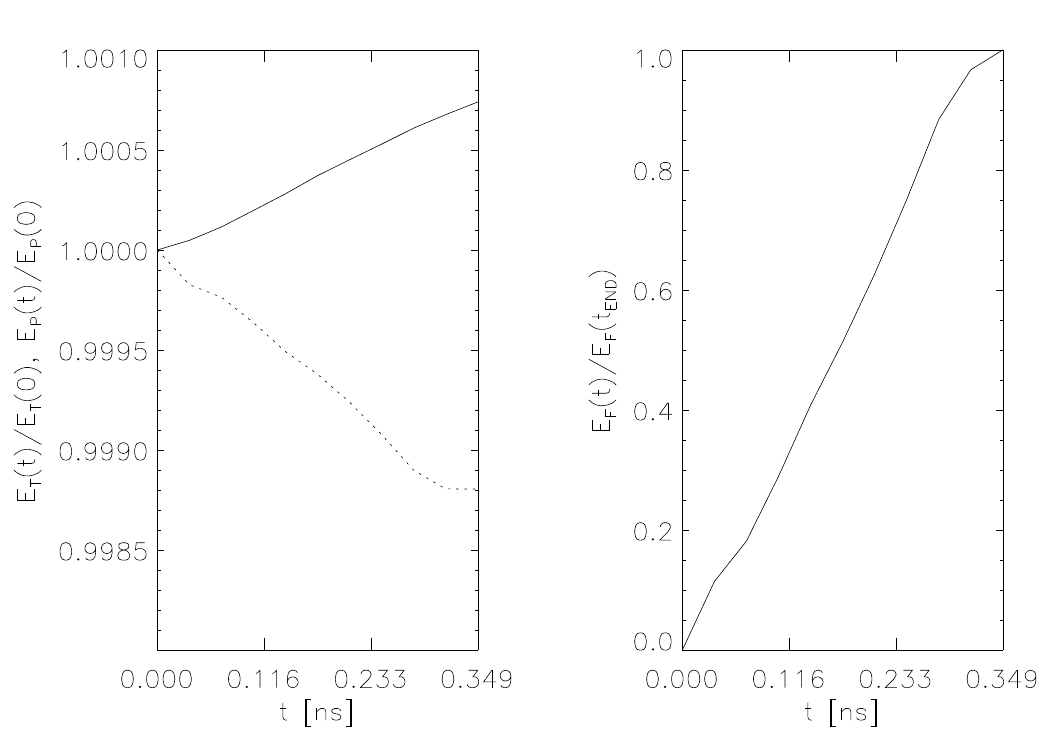}
\end{center}
\caption{ As in Fig.\ref{fig4} but for the case of negative density gradient,
according to equation \ref{e2}.}
\label{fig12}
\end{figure}

The simulation is carried out using EPOCH, a
fully electromagnetic (EM), relativistic PIC code \cite{a15}.
EPOCH is freely available for download from
\url{https://cfsa-pmw.warwick.ac.uk}.
In total seven numerical runs were carried out.
Three runs with uniform, increasing (positive)
and negative (decreasing) density gradients in three and two spatial dimensions (3D and
2D),
and one run with uniform density in one spatial dimension (1D).
Only 3D results are presented here.
See supplemental material at 
\url{http://ph.qmul.ac.uk/~tsiklauri/pwfa1}
for the seven input parameter files used in the simulation.
The simulation parameters are similar to
SLAC's FACET experiment \cite{facet}  and to Ref.\cite{l14}.
In uniform density runs plasma number density is set to
$n_e=n_i=n_0=5\times10^{22}$ m$^{-3}$.    
As described in Ref.\cite{m99}, plasma source in such 
experiments as FACET is produced by photo-ionization
of Lithium vapor contained in heat-pipe oven.
Here similar parameters are used: Lithium plasma temperature of
$T=2.5\times10^{4}$ K and mass ratio of 
$m_i/m_e=12853.1$. Both electron bunch temperatures are also set 
to $T_b=2.5\times10^{4}$ K, producing a very narrow energy/velocity 
distribution spread (see the two bottom panels in Fig.\ref{fig3}).
The simulations domain is split into
$n_x \times n_y \times n_z=7992 \times 24 \times 24$
grid cells in x-, y- and z-directions, respectively.
The actual simulated domain size is within following bounds
$0 \leq x_{max} \leq 10.529748$ cm and 
$0 \leq y_{max},z_{max} \leq 234.229$ $\mu$m. 
This implies the unit grid size in x-direction is
$270 \lambda_D$, while in y- and z-directions
grid size is $200 \lambda_D$. Here
$\lambda_D = v_{th,e}/\omega_{pe}=0.048798$ $\mu$m
is the Debye length, with
$v_{th,e}/c=\sqrt{k_B T/m_e}/c=0.00205332$ being electron thermal
speed and $\omega_{pe}=1.2614673\times10^{13}$ Hz rad
is the plasma frequency.
It maybe counter-intuitive that simulations
with under-resolved Debye length are valid.
The validity is two-fold: (i) the typical total energy error
in our simulations (see e.g. Fig.\ref{fig4}) is 0.00065;
(ii) the experimentally validated simulation results
of Ref.\cite{l14} use $512^3$ grids,
which means 
their grid sizes in y- and z-directions
are under-resolved by a factor of $4918/512=9.6$
(this is because $4918 \lambda_D$ fit into their
y- and z-direction domain sizes of 240 $\mu$m).
Thus it is acceptable to under-resolve Debye length
in such simulations as the electron beams are in blow out regime.
This is understandable, because in plasma PIC
simulation unscreened electric fields, 
within the under-resolved Debye sphere,
lead to onset
of numerical instabilities that result in 
what is known as "numerical heating".
The latter manifests itself through the total energy
increase. On contrary, because finite differencing always leads
to a numerical diffusion, the total energy
must decrease in time. When it increases in time this means
that numerical instability is triggered. However, if 
the total energy error
(see e.g. Fig.\ref{fig4}) is 0.00065 (i.e. 0.065 percent)
such simulations are acceptable. 
The possible reason why under-resolving of Debye length by
factor of 200 in the transverse direction is acceptable,
is because resolving $\lambda_D$ is usually needed to
correctly treat collective plasma effects such as
plasma oscillation. For the plasma wake field acceleration
the relevant spatial scale is electron inertial length
$c/\omega_{pe}$. For the parameters considered 
$c/\omega_{pe}=23.765183$ $\mu$m which means that this
spatial scale is resolved with $23.765/(200\times
0.0488)=2.4$ grid points (recall that $\lambda_D=0.048798$ $\mu$m).
This is not ideal, but since the energy error is small the results are valid, as
typically acceptable level of error is $\approx 0.1-1\%$.

In the positive density gradient runs, 
presented in figures \ref{fig5}, \ref{fig6}, \ref{fig7} and \ref{fig8},
plasma (both background electrons and ions)
number densities vary with distance $x$, in meters, as:
\begin{eqnarray}
n_{PG}(x)=n_0\left(1.0+\frac{9.0 x}{x_{max}}\right)
\biggl[\tanh\left(\frac{x}{0.005x_{max}}\right)+ 
\tanh\left(-\frac{x-x_{max}}{0.005x_{max}}\right)-1.0\biggr].
\label{e1}
\end{eqnarray}
This implies that the density rises from zero to $n_0$ over a length of 
1 mm, then keeps linearly rising to $10 n_0$ and in the
final 1 mm of the domain it falls to zero again.
In equations \ref{e1}-\ref{e4} the distances are quoted in meters.

In the negative density gradient runs, presented in 
figures \ref{fig9}, \ref{fig10}, \ref{fig11} and \ref{fig12}, 
plasma number densities
vary with distance $x$ as:
\begin{eqnarray}
n_{NG}(x)=n_0\left(1.0-\frac{0.9 x}{x_{max}}\right)
\biggl[\tanh\left(\frac{x}{0.005x_{max}}\right)+ 
\tanh\left(-\frac{x-x_{max}}{0.005x_{max}}\right)-1.0\biggr].
\label{e2}
\end{eqnarray}
This implies that the density rises from zero to $n_0$ over a length of 
1 mm, then keeps linearly decreasing to $0.1 n_0$ and in the
final 1 mm of the domain it falls to zero again.
Such background density profiles (equations \ref{e1} and \ref{e2}) allow to use
periodic boundary conditions, which are used
in all numerical simulations presented.

The trailing and driving electron {\it bunches} have the number
densities as follows:
\begin{eqnarray}
n_{T}(x)=n_0 \exp\left[-\frac{(x-10.0c/\omega_{pe})^2}
{2.0(2.0c/\omega_{pe})^2} \right] 
\exp\left[-\frac{(y-y_{max}/2.0)^2}{2.0(c/\omega_{pe})^2}\right]
\exp\left[-\frac{(z-z_{max}/2.0)^2}{2.0(c/\omega_{pe})^2}\right]
\label{e3},
\end{eqnarray}
\begin{eqnarray}
n_{D}(x)=2.5 n_0 \exp\left[-\frac{(x-15.7c/\omega_{pe})^2}
{2.0(c/\omega_{pe})^2} \right] 
\exp\left[-\frac{(y-y_{max}/2.0)^2}{2.0(c/\omega_{pe})^2}\right]
\exp\left[-\frac{(z-z_{max}/2.0)^2}{2.0(c/\omega_{pe})^2}\right].
\label{e4}
\end{eqnarray}
These expressions imply that trailing bunch is centered
on $10.0c/\omega_{pe}=237.65183$ $\mu$m,
has x-length of $\sigma_x=2.0c/\omega_{pe}=47.530365$ $\mu$m,
while 
driving bunch is 2.5 denser than both the background
and trailing bunch, is centered
on $15.7c/\omega_{pe}=373.11337$ $\mu$m and
has x-length of $\sigma_x=c/\omega_{pe}=23.765183$ $\mu$m.
The distance between the trailing and driving
bunches is $5.7c/\omega_{pe}=135.46154$ $\mu$m.
This is the crucial parameter because 
the typical measured and simulated
longitudinal scale of the plasma
wake field for the density of 
$n_0=5\times10^{22}$ m$^{-3}$ is about 
$200$ $\mu$m \cite{l14}. Thus the trailing bunch
should lag behind the driving bunch by
lesser than this length. Our bunch
separation of $\approx 135$ $\mu$m sits
comfortably within this range.
Both electron bunches have
y- and z-lengths of $\sigma_{y,z}=c/\omega_{pe}=23.765183$ $\mu$m
and are centered on $y_{max}/2=
z_{max}/2=117.11432$ $\mu$m.

Both electron bunch initial momenta are set to
$p_x=p_0=1.087587\times10^{-17}$ kg m s$^{-1}$ 
(note that $p_x/(m_e c)=39825.1$),
which corresponds to an initial
energy of $E_0=20.35$ GeV.
There are four plasma species
present in all numerical simulations.
In the 3D runs there are 
$73,654,272$ particles for each of the four species
i.e. roughly $3\times 10^8$ particles
in total.
In the 2D runs there are 
$34,525,440$ particles for each of the four species,
while in 1D each species
have 
$1,438,560$ particles.
The three dimensional runs
take about 27 hours on 192 computing cores, using
Intel Xeon E5-2650 16-core 2.6GHz CPUs with
64 Gb of random access memory (RAM). These use $n_x \times n_y\times n_z=
12\times 4 \times 4 $ domain decomposition
for optimal code performance,
particularly alleviating the particle
dynamical load balancing overheads.
The bottle neck for this type of simulations
is the amount of RAM on each of the 16-core
nodes. With twelve nodes the 
total of $12\times 64=768$ Gb of RAM was available.
Background electrons and ions are
distributed evenly over 192 cores, but
driving and trailing bunches are very localized
and move with nearly at the speed of light. This
puts tight limitation on number of particles
that can be used to resolve bunch electrons
because both bunches must fit computationally
on the 64 Gb RAM compute nodes they fly
through.   

Ref.\cite{l14} used QuickPIC \cite{h6} fully relativistic, 
3D, particle-in-cell model bespoke for simulating 
plasma and laser wake field acceleration. 
QuickPIC is based on the quasi-static approximation, 
which reduces a fully 3D, EM 
field solve and particle push into two spatial dimensions. 
This is achieved by calculating the 
plasma wake based on the assumption of drive beam and/or 
laser not evolving during the time it takes for it 
to pass a plasma particle. The complete EM
fields of the plasma wake and its associated index of 
refraction are then used to evolve the drive beam 
and/or laser using large time steps. Ref.\cite{h6} suggests that 
the algorithm 
reduces the computation time by 2 to 3 orders of magnitude 
without loss of accuracy. In our numerical work the
EPOCH code \cite{a15} is used which is free from simplifying
assumptions of QuickPIC, i.e. it is an explicit,
fully electromagnetic, relativistic PIC code.
Thus, it is interesting to compare the findings of both approaches.

\subsection{Uniform density case}

Fig.\ref{fig1} top row shows contour plots of electric field $E_x$ component at three
times. It can be seen that the yellow half-ellipse, representing positive 
$E_x \approx 2.2 \times 10^{10}$ V/m plasma
wake remains nearly constant in shape and its amplitude and travelled correct distances of
$0.0349\times 10^{-9} \times c=10463$ $\mu$m, 
$0.1744\times 10^{-9} \times c=52283$ $\mu$m and
$0.3488\times 10^{-9} \times c=104567$ $\mu$m within the corresponding times.
Also it can be seen that negative, $E_x \approx -2.5 \times 10^{10}$ V/m, the dark blue half-ellipse,
closely follows the yellow one. It is this region of negative $E_x$ that accelerates
the trailing bunch.
Ref.\cite{l14} has not presented the transverse EM field dynamics. This 
is shown 
in the middle and lower rows of Fig.\ref{fig1}.
In Fig.\ref{fig1} middle row shows contour plots of electric field $E_y$ component at three
times. It can be seen that $E_y$ has quadrupolar structure with two positive peaks on the right side
and two negative dips on the left side of the drive bunch path.
The amplitudes of the quadrupolar $E_y \approx \pm 5 \times 10^{10}$ V/m and are
co-located (across x-position) with $E_x$ positive and negative electrostatic wakes.
Fig.\ref{fig1} lower row shows contour plots of electric field $E_z$ component at three
times. $E_z$ fields show two localized dips with amplitudes of 
$E_z \approx - 5 \times 10^{10}$ V/m that are also spatially co-located with $E_x$ positive 
and negative electrostatic wakes. Note that in the bottom row of Fig.\ref{fig1}
background around the electron bunches appears yellowish color instead of green
that corresponds to zero level. This is
because of the presence of  noise, which in principle can be averaged out
to make background green (zero), but this averaging distorts the shape of
the two dips in $E_z$ and hence the averaging was not performed.
 
Fig.\ref{fig2} shows contour plots of logarithm of electron (top row)  
and ion (bottom row) number densities in (x,y) plane (cut through $z=z_{max}/2$) 
at different time instants. Here the electron number density
includes contributions from background electrons, trailing and driving
electron bunches (i.e. all electrons in the simulation).
It can be seen that the transverse electric field
of the driving bunch expels electrons creating
two density cavities with about 50 $\mu$m size in y-direction and
130 $\mu$m size in x-direction. The latter distance is
estimated as roughly
(1/7)th of distance between $x=1.12$ $\mu$m and $x=1.03$ $\mu$m
i.e. 
$(1.12-1.03)\times 10^4/7 \approx 130$ $\mu$m.
This is consistent with the previous results of Ref.\cite{l14}.
The novelty is, however, in the inclusion of mobile ions.
It can be seen from bottom row panels in Fig.\ref{fig2} that
ion density perturbation can be as large as few times the initial ion density.
The ion density increase has the size of 2000 grids which is about 26351 $\mu$m. 

Fig.\ref{fig3} quantifies the details of background electron, ion, trailing and driving electron bunch
distribution functions at different times:
open diamonds correspond to $t=0$, while blue and red curves to the 
half and the final simulations times, respectively.
It can be gathered from the plot that the background electrons
develop non-thermal tails in the direction of
motion of the trailing and driving electron bunches (i.e. positive
x-direction) with values attaining $p_x\approx 3-4$ $m_e c$.
Ions seem to be heated, rather than developing non-thermal tails, 
which can be witnessed by symmetric broadening of 
the distribution function.
Lower left panel of Fig.\ref{fig3} demonstrates that by end of simulation
time the trailing bunch gains energy of 22.4 GeV, starting from initial 20.4 GeV,
with an energy transfer efficiencies of 65 percent.
The latter is quantified by calculating: 
(i) The
trailing bunch acceleration efficiency
\begin{equation}
AE(t)=
\frac{\int_{p_x > p_0}^\infty      
f_{\rm e, TRAILING}(p_x,t)dp_x}
 { \int_{-\infty}^\infty f_{\rm e, TRAILING}(p_x,t=0) dp_x},
\label{e5}
\end{equation}
where
$p_0=1.087587\times10^{-17}$ kg m s$^{-1}$ 
which corresponds to an initial
energy of $E_0=20.35$ GeV. Note that $p_0/(m_e c)=39825.1$.
Using IDL's 
\verb;int_tabulated; 
built-in function that performs
five-point Newton-Cotes integration with $10^8$ grid points in the $p_x$, the values for 
data of uniform density 3D run are:
$AE(t=0.1744{\rm \; ns})=0.6677$,
$AE(t=0.3488{\rm \; ns})=0.6674$.
All numerical runs use $10^8$ grid points in the $p_x$, thus,
we are confident that indexes $AE(t)$ and $TE(t)$ are calculated
accurately.
(ii) The energy transfer efficiency from the driving bunch
to trailing bunch 
\begin{equation}
TE(t)=
\frac{\int_{p_x > p_0}^\infty      
f_{\rm e, TRAILING}(p_x,t)dp_x}
 { \int_{-\infty}^{p_0} f_{\rm e, DRIVING}(p_x,t) dp_x}.
 \label{e6}
\end{equation}
The values for 
data of uniform density 3D run are:
$TE(t=0.1744{\rm \; ns})=0.6476$,
$TE(t=0.3488{\rm \; ns})=0.6463$.
Note that the physical meaning of $AE(t)$ is the fraction
of trailing bunch electrons with momenta greater that $p_0$
(or with energies greater that $E_0=20.35$ GeV) of
the total number of trailing bunch electrons at $t=0$.
The physical meaning of $TE(t)$ is the fraction
of trailing bunch electrons with momenta greater that $p_0$
(or with energies greater that $E_0=20.35$ GeV) of
the number of driving bunch electrons with momenta less than $p_0$ at time $t$
(not at $t=0$ -- compare the denominators of Equations \ref{e5} and \ref{e6} and note
the different times $t$ used).
Because the energy of accelerated trailing bunch electrons
comes from the deceleration of driving bunch
electrons, the both indexes $AE(t)$ and $TE(t)$ have similar 
values of 65 percent.
Equations \ref{e5} and \ref{e6} contain
infinite integration bounds $\pm \infty$. The employed discretized
version has instead bounds of 
$p_{x, {\rm MAX, MIN}}=\pm 1.4 \times 10^{-17}$ kg m s$^{-1}$.
The range was split into 100 million points. Thus calculation
of $AE(t)$ and $TE(t)$ indexes is accurate.
Lower right panel of Fig.\ref{fig3} demonstrates that by the end of simulation
time starting from initial 20.4 GeV,
the driving bunch loses energy to about 18.3 GeV.

Left panel of Fig.\ref{fig4} shows the behavior of
the total (particles plus EM fields) and 
particle energies, normalized to initial values, respectively. 
It can be seen that the total energy increases due to numerical
heating, but stays within a tolerable value of 0.065 percent.
The particle energy decreases by 0.1 percent.
The right panel shows
EM field energy normalized to its final simulation time
value. In can be seen that it steadily increases as the plasma wake is generated.

\subsection{Positive density gradient case}

Next,  the effect of positive density gradient on PWFA is investigated. 
Fig.\ref{fig5} is similar to Fig.\ref{fig1}, but 
for the case of positive density gradient,
according to equation \ref{e1}.
It can be gathered from top row of Fig.\ref{fig5} that electrostatic plasma
wake becomes spatially localized and, compared to the homogeneous density
3D case, $E_x$ now attains three times larger values of $\pm 6.5 \times 10^{10}$ V/m.
This can be explained by the fact that plasma wake size is prescribed by
the electron inertial length, $c/\omega_{pe}$. Hence, because of the scaling law
$\omega_{pe} \propto \sqrt{n_e}$, the progressively increasing density, into which
driving bunch plows through, creates more localized and stronger wake.
Also more oscillations (peaks and troughs) in the wake can be seen. This is because
the plasma wake is oscillating spatially at $\omega_{pe}$, which is increasing 
with the increase of number density.

Fig.\ref{fig6} is similar to Fig.\ref{fig2} but for the case of positive density gradient numerical simulation run,
which uses density gradient prescribed by equation \ref{e1}. 
It can be seen that, in this case,
electron density cavities are more compact too, commensurate with a more compact
electrostatic wake. The blue strip in the rightmost panel of top row in Fig.\ref{fig6}
is due to background density falling off to zero, as prescribed by
the initial conditions. In the ion density (bottom row)
because x-axis spans entire domain, background density ten-fold increase is 
clearly seen. Ion density perturbation is less profound because it is now
on top of the background density ten-fold increase (due to the 
background density gradient).

Fig.\ref{fig7} is as in Fig.\ref{fig3} but for the case of positive density gradient,
according to equation \ref{e1}. The top row shows that
background electron positive momenta tail is now {\it intermittently} more energetic,
compared to the uniform density 3D case and also
ion heating is  intermittently more stronger.
Bottom left panel shows that trailing bunch energies reach as high as
24.4 GeV. The acceleration  and energy transfer efficiencies in this case
are calculated as follows:
$AE(t=0.1744{\rm \; ns})=0.6100$,
$AE(t=0.3488{\rm \; ns})=0.7326$ and 
$TE(t=0.1744{\rm \; ns})=0.6052$,
$TE(t=0.3488{\rm \; ns})=0.7545$, indicating about 75 percent acceleration  and energy transfer efficiency
by the end simulation time.
The bottom right panel shows that driving bunch partly decelerates and partly
accelerates, i.e. red curve now spreads beyond $E_0$. This can be attributed to
re-acceleration of driving bunch due to the positive density gradient.
Ref.\cite{pt14} 
established that in the case of increasing density along the path of an electron 
beam wave-particle resonant interaction of Langmuir waves 
with the beam electrons leads to an efficient particle acceleration.  
This is because Langmuir waves drift to smaller wavenumbers $k$, allowing them to 
increase their phase speed, $V_{ph}=\omega/k$, and, thus, 
being absorbed by faster electrons. Because the simulations are in the blow out
regime, driving beam number density is 2.5 times denser than
background electron density, Langmuir waves will grow via bump-on-tail
instability on the time scale of $\approx 1/2.5=0.4 \omega_{pe}^{-1}$.
The final simulation time of 0.3488 ns corresponds to $4400 \omega_{pe}^{-1}$.
Thus there will be a plenty of time for the above described effect 
of re-acceleration of driving bunch due to the positive density gradient \cite{pt14} 
to take place.

Fig.\ref{fig8} is as in Fig.\ref{fig4} but for the case of positive density gradient,
according to equation \ref{e1}. It can be seen that now total energy error is
just under 1 percent. This is larger than in the uniform density 3D case, but possibly still
tolerable.

\subsection{Negative density gradient case}

Next,  the effect of negative density gradient on PWFA is investigated. 
Fig.\ref{fig9} is similar to Fig.\ref{fig1}, but 
for the case of negative density gradient,
according to equation \ref{e2}.
It can be seen in top row of Fig.\ref{fig9} that electrostatic plasma
wake becomes spatially wider spread, compared to the homogeneous density
3D case, with larger distances between
the peaks and troughs in x-direction. In the transverse y-direction
size of the wake grows too, such that by the end of simulation
the wake is larger than  $y_{max}= 234.229$ $\mu$m
(note that  data up to 23-rd grid point is plotted, cutting off plot at 224 $\mu$m).
$E_x$ now attains two-and-a-half times smaller value of $-1.1 \times 10^{10}$ V/m.
Again, this can be explained by the fact that plasma wake size is prescribed by
the electron inertial length, $ c /\omega_{pe} \propto 1/ \sqrt{n_e}$. Thus, 
progressively decreasing density 
 creates less localized, wider-spread and thus weaker plasma wake.
Also lesser number of oscillations (peaks and troughs) in the wake is seen. This is because
the plasma wake is oscillating spatially at $\omega_{pe}$, which is decreasing 
with the decrease of number density. Surprisingly, transverse
EM fields (middle and bottom rows in Fig.\ref{fig9}) are similar in their structure
to that of homogeneous density case (middle and bottom rows in Fig.\ref{fig1}).
On the contrary, in positive density gradient case (middle and bottom rows in Fig.\ref{fig5})
more localized $E_y$ and $E_z$ is seen. Thus, it is concluded
that negative gradient affects only electrostatic, $E_x$ 
component and not $E_y$ or $E_z$.

Fig.\ref{fig10} is as in Fig.\ref{fig2} but for the case of negative density gradient,
according to equation \ref{e2}. It can be seen that, in this case,
electron density cavities are larger in size, commensurate with a more wider-spread
electrostatic wake. In the ion density (bottom row)
because x-axis spans entire domain, background density ten-fold decrease is 
clearly seen.
This decrease is prescribed by equation \ref{e2}.
Ion density perturbation is of the same order as the background density.

Fig.\ref{fig11} is as in Fig.\ref{fig3} but for the case of negative density gradient,
according to equation \ref{e2}. The top row shows that
background electron positive momenta tail is now more energetic
compared to the uniform density 3D case and also
ion heating is more stronger.
The bottom left panel shows that trailing bunch energies reach only 
21.4 GeV. The acceleration  and energy transfer efficiencies in this case
are calculated as follows:
$AE(t=0.1744{\rm \; ns})=0.6187$,
$AE(t=0.3488{\rm \; ns})=0.4617$ and 
$TE(t=0.1744{\rm \; ns})=0.6004$,
$TE(t=0.3488{\rm \; ns})=0.4465$, indicating 45 percent acceleration  and energy transfer efficiency
by the end simulation time.
The bottom right panel shows that driving bunch only decelerates 
 i.e. blue and red curves shifts to the left of $E_0$. 
 
Fig.\ref{fig12} is as in Fig.\ref{fig4} but for the case of negative density gradient,
according to equation \ref{e2}. It can be seen that now total energy error is
0.065 percent, which is similar to the uniform density 3D case.

\section{conclusions}

3D, 2D and 1D, particle-in-cell, fully electromagnetic simulations of 
electron plasma wake field acceleration in the blow out regime have been 
carried out. For brevity only 3D results are presented here.
Our aim was to extend earlier results of Ref.\cite{l14} by  
(i) studying the effect of longitudinal density gradient in the light of the results of
Ref.\cite{pt14};
(ii) avoiding use of co-moving simulation box because
the density gradient cases require considering a long domain that fits the entire
gradient; (iii) 
inclusion of ion motion; and (iv) studying fully 
electromagnetic plasma wake fields without quasi-static approximation of QuickPIC \cite{h6}.
It has been shown  
that injecting driving and trailing electron bunches
into a positive density gradient of ten-fold increasing density over 10 cm long plasma,
results in spatially more compact and three times 
larger electric fields ($-6.4 \times 10^{10}$ V/m), 
leading to acceleration of the trailing bunch 
up to 24.4 GeV (starting from initial 20.4 GeV),
with an energy transfer efficiencies from leading to trailing bunch of 75 percent
compared to the
uniform density case.
In the latter case $-2.5 \times 10^{10}$ V/m wave is created
leading to acceleration of the trailing bunch up to 22.4 GeV,
with an energy transfer efficiencies of 65 percent.
It has been shown that injecting the electron bunches
into a negative density gradient of ten-fold decreasing density over 10 cm plasma,
yields spatially more spread wake and two-and-half 
smaller electric fields ($-1.0 \times 10^{10}$ V/m), 
leading to a weaker acceleration of the trailing bunch 
up to 21.4 GeV (starting from initial 20.4 GeV),
with an energy transfer efficiencies from leading to trailing bunch of 45 percent.
It was also shown (not included here) that 2D simulation results are substantially different from
the 3D ones, showing only 10 percent efficiency of trailing bunch acceleration,
while in 1D case understandably no acceleration is seen.
This is because in 1D 
no transverse electric field can be sustained (because transverse directions 
are ignorable) and thus
the driving bunch cannot expel electrons to create
density cavities. 
The ion motions have been also included into consideration, 
showing that in the plasma wake ion number density can
increase over few times the background value. 
Finally, it has been also shown that transverse 
electromagnetic fields in plasma wake are of the same
order as the longitudinal (electrostatic) ones.

In terms of an experimental implementation of the 
proposed PWFA experiments with the
longitudinal density gradient, it is to be remarked that:
(i) The positive density gradient maybe created by driving a hollow piston
in the photo-ionized Lithium vapor plasma contained in heat-pipe oven
\cite{m99}, as in FACET \cite{facet}.
The ten-fold compression of the plasma density may be created by the
piston moving in the same direction as the driving 
and trailing electron bunches that
would fly through a hole in the piston.
(ii) The negative density gradient, e.g. a ten-fold rarefaction, 
maybe created by pulling the hollow piston
the opposite direction to the bunch's motion through 
a hole in the piston.

{\bf Competing interests statement.} Author has no competing interests.

{\bf Funding}. Author was financially supported by 
Leverhulme Trust Research Project Grant RPG-311.

{\bf Acknowledgments.}
Author would like to thank two anonymous referees for careful reading of
the manuscript and useful suggestions.
This research utilized Queen Mary University of London's (QMUL) 
MidPlus computational facilities,       
supported by QMUL Research-IT and funded by UK EPSRC grant EP/K000128/1.
EPOCH code development work 
was in part funded by the UK EPSRC grants 
EP/G054950/1, EP/G056803/1, EP/G055165/1 and EP/ M022463/1 to which Author
has no connection.


\end{document}